\documentclass[journal,10pt]{IEEEtran}
\usepackage{cite}
\usepackage{citesort}
\usepackage{graphicx}
\usepackage[dvips]{color}
\usepackage[dvips]{color}

\usepackage{epsfig}
\usepackage{amsfonts}
\usepackage{amsmath}
\usepackage{bm}
\usepackage{amssymb}
\usepackage{mathrsfs}
\usepackage{stmaryrd}
\usepackage{enumerate}
\usepackage{paralist}

\usepackage{threeparttable}
\usepackage{dcolumn}
\usepackage{multirow}
\usepackage{booktabs}
\usepackage{float}
\usepackage{stfloats}

\usepackage{subfigure}
\usepackage{geometry}
\usepackage{amsthm}

\begin{document}

\title{NOMA Based Calibration for Large-Scale Spaceborne Antenna Arrays}

\author{Yujie Lin, Shuai Wang, Xiangyuan Bu, Chengwen Xing, and Jianping An
\thanks{Manuscript received XXX, XX, 2017; revised XXX, XX, 2017. This work was supported by the National Natural Science Foundation of China under contract U163610110.}

\thanks{The authors are with School of Information and Electronics, Beijing Institute of Technology, Beijing 100081, China (e-mail: yujielin@ieee.org; swang@bit.edu.cn; bxy@bit.edu.cn; chengwenxing@bit.eud.cn; an@bit.edu. cn).}
}

\markboth{IEEE Transactions on Vehicular Technology,~Vol.~XX, No.~XX, XXX~2017}
{}

\maketitle

\begin{abstract}
In the parallel calibration for transmitting phased arrays, the calibration receiver must separate the signals belonging to different antenna elements to avoid mutual interference. Existing algorithms encode different antenna elements' radiation with orthogonal signature codes, but these algorithms are far from desired for large-scale spaceborne antenna arrays. Considering the strictly limited resources on satellites, to improve hardware efficiency of large-scale spaceborne antenna arrays, in this work inspired by the idea of  non-orthogonal multiple access (NOMA) we design a series of non-orthogonal signature codes for different antenna elements by Cyclically Shifting an m-Sequence (CSmS) with different offsets named as CSmS-NOMA signaling. This design can strike an elegant balance between the performance and complexity and is very suitable for large-scale spaceborne antenna arrays. It is shown that no matter how many antenna elements there are to be calibrated simultaneously, CSmS-NOMA signaling needs only one calibrating waveform generator and one matched filter. Hence it is much more efficient than the existing fully orthogonal schemes. In order to evaluate the achievable calibration accuracy, a unified theoretical framework is developed based on which  the relationship between calibration accuracy and signal to noise ratio (SNR) has been clearly revealed. Furthermore, a hardware experiment platform is also built to assess the theoretical work. For all the considered scenarios, it can be concluded that the theoretical, simulated and experimental results coincide with each other perfectly.
\end{abstract}


\section{Introduction}
Large-scale spaceborne phased arrays  are one of the most important enabling techniques for future high-capacity satellite communications [1]--[3]. To deploy high gain beams over targeting areas while maintaining sufficient attenuation elsewhere\cite{Shin2013}, the amplitude \& phase relationship among the array elements must be under full control of the beamformer. For that purpose, the inherent element imbalance, \textrm{a.k.a.} the array channel mismatch, has to be calibrated in advance\cite{PhasedArrayAntennaHandbook}. As a main cause of channel mismatch, mechanical distortion such as elemental antennas displacement or array panel warpage may occur as a result of the tremendous launch impact. Other uncertainties include the drastic temperature fluctuations in the outer space, as well as the aging effect of electronic components. All these factors suggest a pre-launch, on-ground calibration alone cannot solve the problem once and for all, and hence necessitate in-orbit, regular calibrations\cite{Keizer2011},\cite{Brautigam2009}.
\setlength{\parskip}{0\baselineskip}

Practical in-orbit calibration schemes proposed so far have various implementation architectures. In some cases, the calibrating signals are injected into transmitting elements which bypass antennas and loop back to receiving elements, thus the gain and phase drifting of the transceiver channels can be calibrated\cite{Brautigam2009},\cite{WangShuo2011}. The inner calibration architecture is simple and highly integrated, but it leaves the antennas uncalibrated. By contrast, full-path calibration may be realized by placing probing spaceborne antennas  in the near field of the targeted array, so as to establish an outer calibration loop\cite{Takahashi2012},\cite{Takahashi2012_2}. Thus the size, weight and mounting position of the probe becomes a main concern in the design of an outer calibration scheme. With the assistance of the Telemetry \& Telecontrol system, it is technically feasible to move the probing antennas to the ground\cite{SilversteinTSP},\cite{OodoIEICE2001}, leading to another architecture namely the space-ground loop. In this paper we focus on the space-ground loop case, as it possesses the most generality among the above mentioned three. Naturally, the calibration method and the analytical results to be presented are applicable to the inner and outer architectures, too.

Another way of categorizing different calibration schemes is from the viewpoint as for how many elements are calibrated at meantime. Serial approaches deal with one element in a single measurement iteration\cite{Takahashi2012},\cite{Takahashi2012_2}, [12]--[14]. For the sequential measurements to properly reveal element imbalance, they must be finished in a period short enough so that the conditions such as the thermal state of the array channels, the pointing angles of the satellite platform and the atmospheric conditions along the space-ground wireless link remain stationary \cite{SilversteinTSP}. However, this requirement is not always easy to fulfill, as serial calibration schemes, despite their advantage of low complexity, lack efficiency in nature.

Parallel calibration schemes, on the other hand, seem more attractive as they are capable of measuring all the array elements at the meantime \cite{He2014}. Compared with parallel receiving array calibration, parallel transmitting array calibration is more challenging. As the signals radiated by all elements occupy the same frequency and arrive at the same time, the calibration receiver must separate them from one another to avoid mutual interference. An effective approach to this task \mbox{resembles the} code division multiple access (CDMA) technology [16]--[18], as it modulates the signals of different antenna elements with orthogonal signatures, for example, the Walsh codes derived from Hadamard matrices. The calibration receiver comprises a group of matched filters (MF), each corresponding to a signature code and the associated antenna element. At the output of the MFs, perfect element separation is guaranteed by the orthogonality among the signature codes. This orthogonal multiple access (OMA) based concept constitutes the common basis for several independent works about parallel array calibration\cite{Brautigam2009},\cite{SilversteinTSP},\cite{OodoIEICE2001},\cite{WangShuo2009}, and a general mathematical framework can be found in \cite{Besson2010}. However, since each element is assigned with a unique signature code, and the code length must exceeds the number of elements to ensure orthogonality, the complexity invested in generating these calibrating signals at the satellite beamformer and processing them at the ground station is formidable and it could a hardware killer for resource limited satellites, especially for the future large-scale space-borne arrays that comprise hundreds of antenna elements or even more\cite{Han2015}.

In this paper, we will tackle this problem by taking a new look at the parallel transmitting array calibration task from a non-orthogonal multiple access (NOMA) perspective. Recently, NOMA and large-scale MIMO technologies for 5G have attracted intensive interests in the wireless research community\cite{Xing2015},\cite{Ding2014}. On the contrary to conventional medium access protocols such as TDMA, FDMA and CDMA, which allocate different subscribers with orthogonal signatures in the time, frequency or code domain, NOMA deliberately breaks the orthogonality among different subscribers and solve the resultant multi-access interference (MAI) problem by successive interference cancelation (SIC) or message passing algorithms (MPAs). Despite the additional complexity at the receiver for MAI control, many studies have proven NOMA may improve spectral efficiency, boost system connectivity, and meanwhile help cut down the transmission latency and signaling overhead. Inspired by NOMA's success for wireless network, we design novel non-orthogonal signature codes and signal processing techniques that are specially tailored to efficient parallel array calibration. Unlike existing code domain NOMA strategies for 5G which aims at better utilization of resources or enhanced flexibility, the motive of our approach is complexity reduction. Our detailed contributions are listed as follows.
\setlength{\parskip}{-4pt}
\begin{enumerate}
\item In contrast to OMA, our signature codes are generated by cyclically shifting an m-sequence with different offsets. It will be shown that as for this scheme termed as Cyclically Shifted m-Sequence NOMA (CSmS-NOMA), no matter how many elements are to be calibrated simultaneously, it only needs \emph{one} calibrating waveform generator at the beamformer and \emph{one} MF at the receiver, as long as the length of the selected m-sequence is larger than $V$, i.e. the amount of array elements. Hence, CSmS-NOMA based parallel calibration is in general $V$ times more efficient than its OMA counterpart.
\item At the calibration receiver we use a zero-forcing (ZF) equalizer to eliminate the MAI caused by inter-element non-orthogonality. Numerical results shows the ZF-aided CSmS-NOMA performs almost as well as OMA-based strategies in terms of calibration accuracy under various scenarios. Furthermore, in contrast to the MAI controlling methods like SIC or MPA for existing code domain NOMA which requires complexity of the order $\mathcal{O}\left(V^3\right)$ or even more, we have found a peculiar structure for ZF equalizer in the CSmS-NOMA context that allows it to be implemented at a complexity of $\mathcal{O}\left(V\right)$.
\item Closed-form formula are derived to characterize the attainable root mean squared error (RMSE)s of the amplitude and phase calibration results, and the analyses of OMA and CSmS-NOMA based schemes are unified in the same framework. To the best of our knowledge, it is the first time that a clear theoretical relationship among channel mismatch calibration accuracy, signal to noise ratio (SNR), number of elements and length of signature codes, is presented.
\item Besides Monte-Carlo simulations, we have built a hardware prototype system for the performance comparison between OMA and CSmS-NOMA. Extensive numerical examples have shown the analytical, simulated and experimental data coincide with each other perfectly. It means that the proposed theoretical results are applicable in practical hardware platforms in which there are always many nonideal and unpredicted factors limiting the applications of theoretical results.

\end{enumerate}

\setlength{\parskip}{0pt}

The remainder of this paper is organized as follows and here the structure of this paper is given first. In Section II the system model and fundamental concept of OMA-based strategies are presented. In Section III the proposed CSmS-NOMA algorithm is discussed in detail. In Section IV the unified framework for calibration accuracy analysis is proposed. Section V elaborates on the hardware platform. Section VI compares the performance of OMA and CSmS-NOMA based calibration algorithms through theoretical analysis, based on both the computer simulations and hardware experiments. Section VII concludes the paper.

\emph{Notation}: Throughout the paper, without other specifications the light-faced lowercase, bold-faced lowercase and bold-faced uppercase letters are used to denote scalars, vectors and matrices, respectively. The symbols $\textbf{I}_M\in\mathbb{R}^{M\times{M}}$ and $\textbf{O}_{M\times N}\in\mathbb{R}^{M\times{N}}$ stand for the $M$-by-$M$ identity matrix and the ${M\times N}$  all-zero matrix. For vector $\boldsymbol{a}\in\mathbb{R}^{M\times 1}$ or $\boldsymbol{a}\in\mathbb{C}^{M\times 1}$, $\boldsymbol{a}(m)$ denotes its $m$-th entry, and of course $1\leq m\leq M$. Finally, $\text{E}(\cdot)$, $\text{var}(\cdot)$, $(\cdot)^\text{T}$, $(\cdot)^\text{H}$, and $\text{det}(\cdot)$ represent the operators for mean, variance, transpose, conjugate transpose, and determinant, respectively.

\begin{figure*}\centering
\includegraphics[height=60mm]{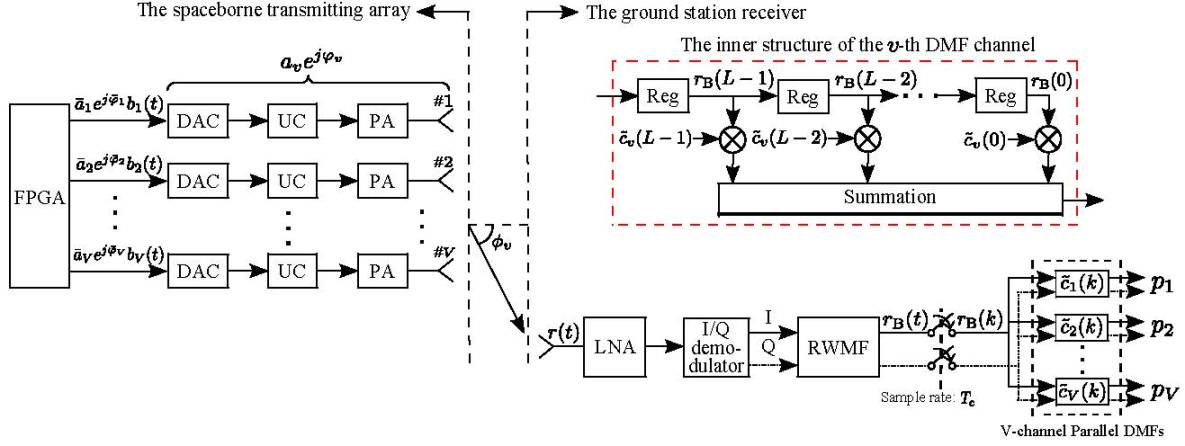}
\caption{Block diagram of the spaceborne transmitting phased array calibration system\label{TransmittingReceiving}.}
\end{figure*}

\section{System Model}\label{S2}

In the upper-left of Fig.1 it is the diagram of a downlink digital beamforming (DBF) array that consists of $V$ elemental antennas. The $V$ digital intermediate frequency (IF) signals are generated by an field programmable gate array (FPGA), and then they go through their respective chains of digital analog converter (DAC), up converter (UC), power amplifier (PA) and then sent from each antenna element. Finally, the signal received at the ground-based calibration probe is of the following mathematical formula
\begin{equation}\label{eq_r_t} 
r(t)=L_p\sum_{v=1}^{V}\bar{a}_va_vb_v(t)\cos{\left[2{\pi}f_\text{c}(t)+\bar{{\varphi}}_v+{\varphi}_v+\phi_v\right]}+n(t),
\end{equation}where $L_p$ the path loss coming from propagation distance.
The complex-valued channel gain (CCG) of channel $v$ is denoted by $a_ve^{j\varphi_v}$, which  reflects the composite effect of DAC, UC, PA and antenna element on channel $v$.
In addition, $\bar{a}_ve^{j\bar{\varphi}_v}$ is the complex-valued weight performed by the DBF on antenna element $v$. The baseband calibrating waveform transmitted by this element is denoted by $b_v(t)$. Finally, $n(t)$ is the white Gaussian noise at the calibration receiver,
The parameters $\{\phi_v|v=1,2,\cdots,V\}$ in the channel gains are determined by the structure of the array as well as the relative position of the array center to the ground-based probe \cite{SilversteinTSP}.

It is worth noting that calibration aims to determine the channel mismatch instead of $\{a_ve^{j\varphi_v}|v=1,2,\cdots,V\}$ themselves. In other words, our focus is their relative values. Taking $a_1e^{j\varphi_1}$ as the reference without loss of generality, the $(V-1)$ pairs of relative values to be determined are
\setcounter{equation}{1}
\begin{equation}
\left\{20\lg\left(\frac{a_v}{a_1}\right),\left(\varphi_v-\varphi_1\right)\Big|v=2,3,\cdots,V\right\}.
\end{equation}In calibration procedure, the DBF may assign uniform weights on all elements, i.e., $\bar{a}_ve^{j\bar{\varphi}_v}=1$. The propagation loss $L_p$ has no influence on the channel mismatch. The values of $\{\phi_v|$ $v=1,2,\cdots,V\}$ are assumed to be known a priori because in practice these parameters can be obtained by telemetry systems \cite{SilversteinTripulse}. Based these observations, we can let $L_p=1$ and $\{\phi_v=0|v=$ $1,2,\cdots,V\}$ in (\ref{eq_r_t}) for mathematical convenience, and thus  a much neater form of (\ref{eq_r_t}) can be achieved
\begin{equation}\label{eq_r_t2}
r(t)=\sum_{v=1}^{V}a_vb_v(t-\tau)\cos{\left[2{\pi}f_\text{c}(t-\tau)+{\varphi}_v\right]}+n(t).
\end{equation}

The calibrating waveforms $\{b_v(t)|v=1,2,\cdots,V\}$ are baseband direct sequence spread spectrum (DSSS) signals, i.e.,
\begin{equation} \label{b_k(t)}
b_v(t)=\sum_{l=0}^{L-1}\boldsymbol{c}_v(l)h(t-lT_c), \quad t\in[0,T_\text{s})
\end{equation}
where $\boldsymbol{c}_v\in\mathbb{R}^{L\times1}$ is a length-$L$ signature code which is assigned uniquely to the $v^{\rm{th}}$ element. Herein we assume the signature codes are bipolar sequences, i.e. $\boldsymbol{c}_v(l)=\frac{\pm1}{\sqrt{L}}$ for all $v$ and $l$, where the factor $\frac{1}{\sqrt{L}}$ makes sure $\boldsymbol{c}_v^\text{T}\boldsymbol{c}_v=1$. In (\ref{b_k(t)}), $h(t)$ is an energy-normalized rectangular pulse as follows
\begin{align} \label{chipwaveform}
h(t)=\left\{
   \begin{array}{ll}
    \frac{1}{T_\text{c}},\quad&\text{if}\;\;t\in[0,T_\text{c}),\\
    \,0,\,&\text{otherwise}.
   \end{array}
 \right.
\end{align}As $\boldsymbol{c}_v$ comprises $L$ entries, the entire waveform $b_v(t)$ has a duration of $T_\text{s}=LT_\text{c}$. In practice, the calibrating waveform is usually transmitted repeatedly, and thus $b_v(t)$ can be regarded as a periodic signal with period $T_\text{s}$.

As shown in the lower-right of Fig. \ref{TransmittingReceiving}, the received signal goes sequentially through a low-noise amplifier (LNA), a quadrature (I/Q) demodulator and a dual-channel rectangular waveform matched filter (RWMF), and finally the received signal is transferred into the following complex-valued baseband signal formula
\begin{equation} \label{r_B(t)}
r_\text{B}(t)=\sum_{v=1}^{V}w_v\sum_{l=0}^{L-1}\boldsymbol{c}_v(l)H(t-lT_c-\tau)+n_\text{B}(t),
\end{equation}
where $\{w_v=a_ve^{j{\varphi}_v}|v=1,2,\cdots,V\}$ and $H(t)=h(t)\ast h(t)$ with $H(t)|_{t=T_\text{c}}=1$. The symbol $n_\text{B}(t)$ denotes the baseband complex-valued noise. Assuming perfect synchronization that the receiver samples $r_\text{B}(t)$ at the epochs of $t=(k+1)\cdot T_\text{c}$, with $k=0,1,2,\cdots$, we obtain the following discrete signal
\begin{equation} \label{r_B(n)}
r_\text{B}(k)=\sum_{v=1}^{V}w_v\boldsymbol{c}_v(k)+n_\text{B}(k),
\end{equation}
where $\boldsymbol{c}_v(k)=\boldsymbol{c}_v\left(\text{mod}(k,L)\right)$ if $k>L$. Note that $n_\text{B}(k)$ is a zero-mean, circularly-symmetric, complex-valued white Gaussian sequence. Specifically, we have $\text{E}[n_\text{B}(k)n^\ast_\text{B}(k')]=0$ for any $k\neq k'$, and $\text{E}\left[|n_\text{B}(k)|^2\right]=\sigma^2$. $r_\text{B}(k)$ is then fed to $V$ parallel channels of digital matched filter (DMF). As shown in the red dash-line square of Fig.1, the unit impulse response (UIR) of the $v^{\rm{th}}$ DMF is
\begin{align} \label{dmf}
\tilde{c}_v(k)=\left\{
   \begin{array}{ll}
    \boldsymbol{c}_v(L-1-k),\quad&\text{if}\;\;0\leq k\leq L-1,\\
    \quad\quad\quad0,&\text{otherwise}.
   \end{array}
 \right.
\end{align}

When the signature codes in $r_\text{B}(k)$ become phase-aligned with the UIRs of the DMFs, the outputs of all the $V$ DMFs will yield correlation peaks simultaneously. To gain more insight, consider a length-$L$ segment of $r_\text{B}(k)$, i.e., $\boldsymbol{r}_\text{B}=[r_\text{B}(0),$ $r_\text{B}(1),\cdots,r_\text{B}(L-1)]^\text{T}\in\mathbb{C}^{L\times1}$. Referring to (\ref{r_B(t)}) and (\ref{dmf}), we may find that correlation peaks will appear when the sample $r_\text{B}(L-1)$ enters the DMF bank, and the peak value yielded by the DMF on the $v^{\rm{th}}$ channel, say $p_v$, is given by
\begin{equation}\label{pv}
p_v=\boldsymbol{c}_v^\text{T}\boldsymbol{r}_\text{B}.
\end{equation}
Now stacking the peak values on all the $V$ channels in a vector $\boldsymbol{p}=[p_1,p_2,\cdots,p_V]^\text{T}$, we have
\begin{equation} \label{peak}
\boldsymbol{p}=\boldsymbol{C}^\text{T}\boldsymbol{r}_\text{B},
\end{equation}
where $\boldsymbol{C}=\left[\boldsymbol{c}_1,\boldsymbol{c}_2,\cdots,\boldsymbol{c}_V\right]\in\mathbb{R}^{L\times V}$. Recalling the definition of $r_\text{B}(k)$ and $\boldsymbol{r}_\text{B}$, $\boldsymbol{r}_\text{B}$ can be further written as
\begin{equation} \label{rB}
\boldsymbol{r}_\text{B}=\boldsymbol{C}{\boldsymbol{w}}+\boldsymbol{n}_\text{B},
\end{equation}
where the vector $\boldsymbol{w}=[w_1,w_2,\cdots,w_V]^\text{T}\in\mathbb{C}^{V\times1}$, while the noise term $\boldsymbol{n}_\text{B}\in\mathbb{C}^{L\times1}$ is a length-$L$ segment of $n_\text{B}(k)$ that satisfies $E\{\boldsymbol{n}_\text{B}\boldsymbol{n}^\text{H}_\text{B}\}=\sigma^2\textbf{I}_L$. Plugging (\ref{rB}) into (\ref{peak}) leads to
\begin{equation} \label{rB2}
\boldsymbol{p}=\boldsymbol{R}{\boldsymbol{w}}+\boldsymbol{C}^\text{T}\boldsymbol{n}_\text{B},
\end{equation}where $\boldsymbol{R}=\boldsymbol{C}^\text{T}\boldsymbol{C}$.

Note the predominant idea behind calibration for antenna arrays is to guarantee the orthogonality between different calibrating waveforms transmitted from different antenna elements. In other words, in nature this problem is exactly a multiple access problem. Up to date, mutually orthogonal signature codes e.g., the Walsh sequences used as the baseband calibrating waveform, which satisfy $\boldsymbol{R}=\textbf{I}_V$ \cite{Brautigam2009},\cite{SilversteinTSP},\cite{OodoIEICE2001},\cite{WangShuo2009},\cite{Besson2010}. For strictly orthogonal signature codes, the scheme is named as orthogonal multiple access (OMA) scheme.
Correspondingly, (\ref{rB2}) can be rewritten as  $\boldsymbol{p}={\boldsymbol{w}}+\boldsymbol{C}^\text{T}\boldsymbol{n}_\text{B}$. As $\boldsymbol{x}=\boldsymbol{C}^\text{T}\boldsymbol{n}_\text{B}$ is zero mean, $\boldsymbol{p}$ is an unbiased estimate of $\boldsymbol{w}$.  A new symbol $\tilde{\boldsymbol{w}}$ is introduced to denote the estimate of $\boldsymbol{w}$,
\begin{equation} \label{wOforthogonalScheme}
\tilde{\boldsymbol{w}}=\boldsymbol{p}={\boldsymbol{w}}+\boldsymbol{x}.
\end{equation}
Interestingly, here $\boldsymbol{x}$ is also a white noise vector with $\text{E}\{\boldsymbol{x}{\boldsymbol{x}}^\text{H}\}=\boldsymbol{C}^\text{T}\text{E}\{\boldsymbol{n}_\text{B}\boldsymbol{n}_\text{B}^\text{H}\}\boldsymbol{C}=\sigma^2\textbf{I}_V$. Given $\tilde{\boldsymbol{w}}=[\tilde{w}_1,\tilde{w}_2,\cdots,\tilde{w}_V]^\text{T}\in\mathbb{C}^{V\times1}$, the estimated mismatch of element $2\thicksim V$ with respect to the reference element (element $1$) is derived to be
\begin{equation} \label{AmpEstimte}
\left\{20\lg\left(\left|\frac{\tilde{w}_v}{\tilde{w}_1}\right|\right),\left(\text{arg}(\tilde{w}_v)-\text{arg}(\tilde{w}_1)\right)\Big|2\leq v\leq V\right\}.
\end{equation}

\section{Parallel Calibration Based on NOMA Technology}\label{S3}

OMA-based designs are incompetent in the case with large-scale antenna arrays. Finding strictly orthogonal code sequences for a large number of antenna elements are very challenging and costly. Nevertheless, it is desirable to design calibrating sequences owning the characteristics of OMA-based designs.
Inspired by the idea of NOMA for future 5G ground-based cellular networks, in the following a novel NOMA based scheme is proposed for the large-scale spaceborne antenna array calibration.

In our work the famous m-sequences are used for NOMA signaling design. Note that both cyclic shifting and ZF works well with arbitrary codes, not only with m-sequences specifically. However, if we choose the seed code for cyclic shifting at will, then the matrix-by-vector product will require much higher complexity.  By constraint, it is well known the periodic auto-correlation of a normalized m-sequence takes only two possible values, namely $1$ for zero offset, or $-\frac{1}{L}$ for any other offsets \cite{ChenHH}. This fact can facilitate the following design.

\subsection{Cyclically Shifted m-Sequence Signaling}

The signature code assigned to the first element is denoted as
$\boldsymbol{m}=[\boldsymbol{m}(0),$ $\boldsymbol{m}(1),\cdots,\boldsymbol{m}(L-1)]^\text{T}$. In addition, $\boldsymbol{m}$ is a normalized m-sequence i.e., $\boldsymbol{m}^\text{T}\boldsymbol{m}=1$. Then the $q$-th cyclically shifted version of $\boldsymbol{m}$ is defined as $
\boldsymbol{m}^q=[\boldsymbol{m}(L-q),\cdots,\boldsymbol{m}(L-1),\boldsymbol{m}(0),\cdots,\boldsymbol{m}(L-q-1)]^\text{T}$,
where $0\leq q\leq L-1$, and by default $\boldsymbol{m}^0=\boldsymbol{m}$. In CSmS-NOMA, the signature code allocated to the $v$-th element is $\boldsymbol{m}^{q_v}$, and for $1\leq v\leq V$ we assume $q_1=0<q_2<q_3<\cdots<q_{V}\leq L-1$, without loss of generality. Recalling (\ref{b_k(t)}), the baseband calibrating waveform of element $v$ can be reformulated as
\begin{equation} \label{m(t)}
b_v(t)=\sum_{l=0}^{L-1}\boldsymbol{m}^{q_v}(l)h(t-lT_c), \quad t\in[\,0,LT_\text{c}).
\end{equation}

As investigated in Section \ref{S2}, $\{b_v(t)|1\leq v\leq V\}$ are all periodic waveforms with equal period  $T_\text{s}=LT_\text{c}$. Moreover, $\boldsymbol{m}^{q_v}$ is the $q_v$-entries cyclically shifted version of $\boldsymbol{m}$. Based on these two facts the following equality holds
\begin{equation} \label{RelationOfmkAndm}
b_v(t)=b_1(t-q_vT_\text{c}),\quad -\infty<t<+\infty.
\end{equation}
It implies $b_v(t)$ can be generated simply via delaying $b_1(t)$ by $q_vT_\text{c}$. For DBF systems, these delays can be inserted flexibly and precisely by FPGAs or DSPs. The delaying strategy is much more convenient than the traditional OMA-based schemes which produce all the $V$ calibrating waveforms independently.

At the CSmS-NOMA calibration receiver, in principle, we may still use the structure as shown in Fig.1, which comprises $V$ parallel DMFs. However, the cyclic relationship among $\{b_v(t)|1\leq v\leq V\}$ of (\ref{RelationOfmkAndm}) provides us with a more efficient alternative based on only one DMF. To see it, let us first examine (\ref{r_B(n)}) in the CSmS-NOMA setup, that is
\begin{equation}\label{r_B(n)2}
r_\text{B}(k)=\underbrace{\sum_{v=1}^{V}w_v\boldsymbol{m}^{q_v}(k)}_{\overline{r}_B(k)}+n_\text{B}(k),
\end{equation}
where $\overline{r}_B(k)$ is the desired signal. Substitute $r_\text{B}(k)$ in (\ref{r_B(n)2}) into a DMF toned to element 1, and thus its UIR can be written in the following form
\begin{equation}  \label{mdmf}
u(k)=\left\{
   \begin{array}{ll}
    \boldsymbol{m}(L-1-k),\quad&\text{if}\;\;0\leq k\leq L-1,\\
    \quad\quad\quad0,&\text{otherwise}.
   \end{array}
 \right.
\end{equation}

\begin{figure}\centering
\centering
\includegraphics[height=50mm]{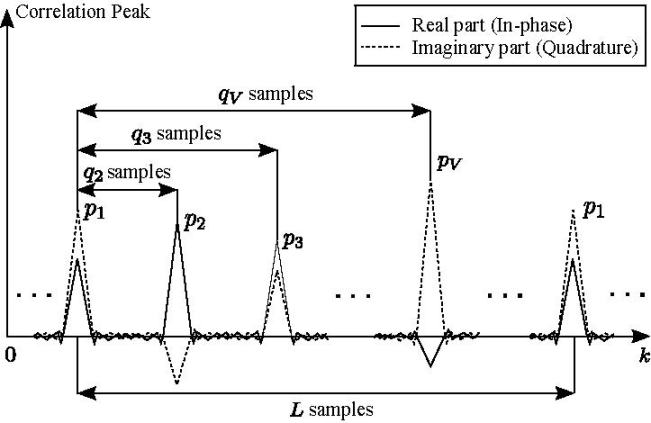}
\caption{The sequential relationships among the correlation speaks of CSmS-NOMA.\label{CorrelationPeak}}
\end{figure}

When $r_\text{B}(L-1)$ enters the DMF, the component in $r_\text{B}(k)$ corresponding to element 1, namely $w_1\boldsymbol{m}(k)$, is matched by the UIR of (\ref{mdmf}), and this will contribute a correlation peak to the output of the filter. Interestingly, $q_2$ samples later, when $r_\text{B}(L-1+q_2)$ enters the DMF, the component corresponding to element 2, i.e., $w_2\boldsymbol{m}^{q_2}(k)$, will be matched, because $\boldsymbol{m}^{q_2}(k)$ is simply a $q_2$-samples delayed version of $\boldsymbol{m}(k)$. Therefore, the correlation peak of element 2 will appear $q_2$ samples later than that of element 1. It is obvious that the correlation peaks pertaining to element $3,4,\cdots,V$ appear consecutively, and their offsets relative to element 1's peak are $q_3,q_4,\cdots,q_V$ samples, respectively, as illustrated in Fig. \ref{CorrelationPeak}. By recording the output of the DMF at appropriate epochs, the receiver is capable to collect the $V$ peak values $\{p_1,p_2,\cdots,p_V\}$ for further processing. In other words, CSmS-NOMA enables the receiver to correlate the $V$ calibrating components sequentially with only one DMF, instead of simultaneously with $V$ DMFs as in the OMA-based schemes. In practice, once the first correlation peak is found, then the following $(V-1)$ ones could be located precisely, since the relative intervals are predefined.

\setcounter{equation}{26}
\begin{figure*}[hb]
\hrulefill
\begin{align}
G_{v,1}^\text{RMSE}&\approx\frac{10}{\ln (10)}\sqrt{\frac{\hat{\sigma}_1^2\hat{\mu}_v^2}
{\hat{\mu}_1^4}+\frac{\hat{\sigma}_v^2}{\hat{\mu}_1^2}
-\frac{2\hat{\rho}_{v,1}\hat{\sigma}_1
\hat{\sigma}_v\hat{\mu}_v}{\hat{\mu}_1^3}
+\left(\frac{\hat{\mu}_v}{\hat{\mu}_1}
+\frac{\hat{\sigma}_1^2\hat{\mu}_v}{\hat{\mu}_1^3}
-\frac{\hat{\rho}_{v,1}\hat{\sigma}_1\hat{\sigma}_v}
{\hat{\mu}_1^2}-1\right)^2}, \label{ApproxAmp} \\
P_{v,1}^\text{RMSE}&\approx\frac{180}{\pi}
\sqrt{\frac{\sigma_v^2}{2a_v^2}+\frac{\sigma_1^2}{2a_1^2}
-\frac{2\rho_{v,1}\sigma_1\sigma_v\cos(\varphi_1-\varphi_v)}
{2a_va_1+\rho_{v,1}\sigma_1\sigma_v\cos(\varphi_1-\varphi_v)}}. \label{ApproxPhs}
\end{align}
\end{figure*}
\setcounter{equation}{18}

\subsection{Inter-Element Interference Elimination}

As discussed above, CSmS-NOMA achieves complexity savings at the cost of breaking the orthogonality among the signature codes. Now we investigate how the non-orthogonal signature codes impact the calibration accuracy and explore the possible countermeasure. To begin with, let us examine the correlation peak of element 1, i.e. $p_1=\boldsymbol{m}^\text{T}\boldsymbol{r}_\text{B}$, where $\boldsymbol{r}_\text{B}=[r_\text{B}(0),r_\text{B}(1),$ $\cdots,r_\text{B}(L-1)]^\text{T}$ $\in\mathbb{C}^{L\times1}$ could be further expressed as $\boldsymbol{r}_\text{B}=\overline{\boldsymbol{r}}_\text{B}+\boldsymbol{n}_\text{B}$, with $\overline{\boldsymbol{r}}_\text{B}$ and $\boldsymbol{n}_\text{B}\in\mathbb{C}^{L\times1}$ being the sampled vectors of the composite calibrating signal and the noise, respectively. By consulting (\ref{r_B(n)2}), it is found that $\overline{\boldsymbol{r}}_\text{B}=\left[\boldsymbol{m},\boldsymbol{m}^{q_2},\cdots,\boldsymbol{m}^{q_V}\right]\boldsymbol{w}$,
where $\boldsymbol{w}\in\mathbb{C}^{V\times1}$ has been defined in (\ref{rB}). In the following we focus on the noise-free part of $p_1$ i.e.,
\begin{equation}\label{overlineP1CSmS-NOMA}
\overline{p}_1=\boldsymbol{m}^\text{T}\overline{\boldsymbol{r}}_\text{B}=\left[\boldsymbol{m}^\text{T}\boldsymbol{m},\boldsymbol{m}^\text{T}\boldsymbol{m}^{q_2},\cdots,\boldsymbol{m}^\text{T}\boldsymbol{m}^{q_V}\right]\boldsymbol{w},
\end{equation}
as the signature codes generated by CSmS-NOMA are no longer orthogonal, there is no guarantee that $\boldsymbol{m}^\text{T}\boldsymbol{m}^{q_2}=$ $\boldsymbol{m}^\text{T}\boldsymbol{m}^{q_3}=\cdots=\boldsymbol{m}^\text{T}\boldsymbol{m}^{q_V}=0$. Therefore, though $\boldsymbol{m}^\text{T}\boldsymbol{m}=1$, we cannot expect $\overline{p}_1=\boldsymbol{w}(1)=$ $w_1$, since the other $(V-1)$ entries of $\boldsymbol{w}$ may contribute unwanted inter-element interference (IEI) to $\overline{p}_1$. When $V$ is large, IEI deviates the calibration results severely from their true values even in a noise-free case.

For large scale antenna arrays, the optimal transceiver to combat IEI is called zero forcing (ZF).
This technique originates from the research on multi-user detection (MUD) for CDMA systems\cite{YangLLbook},\cite{Madhow}. To see how ZF is adapted to the CSmS-NOMA setup,
let us consider all the $V$ correlation peaks collectively by staking their noise-free parts into $\overline{\boldsymbol{p}}=[\overline{p}_1,\overline{p}_2,\cdots,\overline{p}_V]^\text{T}$. More particularly, we have
\begin{align}
\overline{\boldsymbol{p}}&=\underbrace{\left[\begin{array}{cccc}
      \boldsymbol{m}^\text{T}\boldsymbol{m} & \boldsymbol{m}^\text{T}\boldsymbol{m}^{q_2} & \cdots & \boldsymbol{m}^\text{T}\boldsymbol{m}^{q_V} \\
      \boldsymbol{m}^\text{T}\boldsymbol{m}^{-q_2} & \boldsymbol{m}^\text{T}\boldsymbol{m} & \cdots & \boldsymbol{m}^\text{T}\boldsymbol{m}^{q_V-q_2} \\
      \vdots & \vdots & \ddots & \vdots \\
      \boldsymbol{m}^\text{T}\boldsymbol{m}^{-q_V} & \boldsymbol{m}^\text{T}\boldsymbol{m}^{q_2-q_V} & \cdots & \boldsymbol{m}^\text{T}\boldsymbol{m} \\
    \end{array}\right]}_{\boldsymbol{M}}\boldsymbol{w} \nonumber \\
    &=\boldsymbol{M}\boldsymbol{w}.
\end{align}
Note that in the above formula the other $(V-1)$ entries in $\overline{\boldsymbol{p}}$ than $\overline{p}_1$ can be derived following the lines of (\ref{overlineP1CSmS-NOMA}).

IEI elimination amounts to recovering $\boldsymbol{w}$ from $\overline{\boldsymbol{p}}$, and ZF accomplish this in the most direct way, i.e. by multiplying $\overline{\boldsymbol{p}}$ with $\boldsymbol{M}^{-1}$. In practice, the receiver is only able to obtain $\boldsymbol{p}$, which is a noisy version of $\overline{\boldsymbol{p}}$, therefore the ZF-based estimate of $\boldsymbol{w}$ goes like
\begin{equation} \label{DC}
\tilde{\boldsymbol{w}}=\boldsymbol{M}^{-1}\boldsymbol{p},
\end{equation}
where the inverse matrix $\boldsymbol{M}^{-1}$ could be precomputed, because $\boldsymbol{M}\in\mathbb{R}^{V\times V}$ is solely dependent on the CSmS-NOMA parameters, such as the m-sequence $\boldsymbol{m}$ and the $V-1$ offsets $\{q_2,q_3,\cdots,$ $q_V\}$. Once $\tilde{\boldsymbol{w}}$ is found, amplitude and phase mismatch among different elements could be obtained using (\ref{AmpEstimte}). It is well known the periodic auto-correlation of a normalized m-sequence takes only two possible values, i.e., $1$ for zero offset, and $-\frac{1}{L}$ for any other offsets \cite{ChenHH}. This property guarantees a unique structure of $\boldsymbol{M}$ in the following form
\begin{align}
\boldsymbol{M}=\frac{1}{L}\left[\begin{array}{cccc}
      L & -1 & \cdots & -1 \\
      -1 & L & \cdots & -1 \\
      \vdots & \vdots & \ddots & \vdots \\
      -1 & -1 & \cdots & L \\
    \end{array}\right]_{V\times V},
\end{align}
regardless of the specific values of $V$ and $\{q_2,\cdots,q_V\}$. After tedious but straightforward mathematical derivations, the inverse of $\boldsymbol{M}$ is derived to be
\begin{equation}\label{UniqueStructure}
\boldsymbol{M}^{-1}=\left[\begin{array}{cccc}
      b & a & \cdots & a \\
      a & b & \cdots & a \\
      \vdots & \vdots & \ddots & \vdots \\
      a & a & \cdots & b \\
    \end{array}\right]_{V\times V},
\end{equation}
where $a=\frac{L}{(L+1)(L-V+1)}$, $b=\frac{L(L-V+2)}{(L+1)(L-V+1)}$. Based on (\ref{UniqueStructure}), the first entry of $\tilde{\boldsymbol{w}}$ can be computed as $\tilde{w}_1=a\left(\sum_{v'=1}^{V}p_{v'}\right)-p_1(b-a)$. Although the sum herein requires $(2V-2)$ real-valued additions (RAs), yet it can be reused in computing the other $(V-1)$ entries. Besides this, the computation of $\tilde{w}_1$ still needs $4$ more real-valued multiplications (RMs) and $2$ more RAs. Hence invoking (\ref{DC}) to evaluate $\tilde{\boldsymbol{w}}$ calls for $4V$ RMs and $(4V-2)$ RAs in total, that is a lot more efficient than by ZF without using m-sequence.

As an ending note about ZF, we would like to point out that ZF eliminates the IEI at the price of noise amplification, as found in \cite{Madhow}. This observation, interpreted in our context, suggests ZF-aided CSmS-NOMA is in general more vulnerable to noise than OMA.
The noise enlarging effect (NEE) of ZF will be further examined in our numerical study in the following section. Also, we will show that NEE turns out negligible in most cases, provided the code length has been properly selected according to the number of elements.

\section{Calibration Performance Analysis}\label{S4}

In the following analysis, the root mean squared error (RMSE) (in decibel) of the estimated gain mismatch between element  $v\geq2$  and element $1$ is used as calibration accuracy metrics
\begin{equation} \label{AmpRMSE}
G_{v,1}^{\text{RMSE}}=\sqrt{\text{E}\left\{\left[20\log_{10}\left(\left|\frac{\tilde{w}_v}{\tilde{w}_1}\right|\right)-20\log_{10}\left(\frac{a_v}{a_1}\right)\right]^2\right\}},
\end{equation}
and the RMSE (in degree) of the phase mismatch estimate
\begin{equation} \label{PhsRMSE}
P_{v,1}^{\text{RMSE}}=\frac{180}{\pi}\sqrt{\text{E}\left\{\left[\left(\text{arg}(\tilde{w}_v)-\text{arg}(\tilde{w}_1)\right)-\left(\varphi_v-\varphi_1\right)\right]^2\right\}}.
\end{equation}

Given the specific system parameters, such as the SNR, the number of elements and the signature codes used, etc., the performance metrics $G_{v,1}^{\text{RMSE}}$ and $P_{v,1}^{\text{RMSE}}$ could be assessed by simulations \cite{He2014},\cite{WangShuo2009}. Unfortunately, this is far from desired as these results have close relationship with many involved parameters.
To the best of our knowledge, a theoretical accuracy analysis is still largely open in the literature. In the following discourse, we will fill this gap by developing a general framework for computing $G_{v,1}^{\text{RMSE}}$ and $P_{v,1}^{\text{RMSE}}$ approximately, which applies to both OMA-based schemes and CSmS-NOMA-ZF.

\subsection{Theoretical Basis}\label{S4.A}

At the key part of our analytical framework is based on the following proposition.

\emph{Proposition 1.} The estimated CCGs of element $1$ and element $v$ ($2\leq v\leq V$) may be written as $\tilde{w}_1=w_1+x_1$ and $\tilde{w}_v=w_v+x_v$, where $w_1=a_1e^{j\varphi_1}$ and $w_v=a_ve^{j\varphi_v}$ stand for the true values of CCGs. The noise terms $x_1=x_{\text{I},1}+j\cdot x_{\text{Q},1}$ and $x_v=x_{\text{I},v}+j\cdot x_{\text{Q},v}$ are complex-valued Gaussian RVs. Specifically, $x_{\text{I},1}$ and $x_{\text{Q},1}\in$ $\mathbb{R}^{1\times1}$ are independent identically-distributed (i.i.d.) zero-mean Gaussian RVs, with $\text{var}\left(x_{\text{I},1}\right)=\text{var}\left(x_{\text{Q},1}\right)=\sigma_1^2/2$. The same assumption also applies to $x_{\text{I},v}$ and $x_{\text{Q},v}$, except $\text{var}\left(x_{\text{I},v}\right)=$ $\text{var}\left(x_{\text{Q},v}\right)=\sigma_v^2/2$. The correlation coefficient between $x_v$ and $x_1$ is defined as
\begin{equation} \label{rho}
\rho_{v,1}=\frac{\text{E}\left(x_vx_1^\ast\right)}{\sigma_v\sigma_1},
\end{equation}
and $\rho_{v,1}\in\mathbb{R}^{1\times1}$. Based on these characteristics, $G_{v,1}^{\text{RMSE}}$ and $P_{v,1}^{\text{RMSE}}$ could be approximated by the two formulas presented at the bottom of the page, given $a_1^2/\sigma_1^2$ and $a_v^2/\sigma_v^2$ are sufficiently large. In (\ref{ApproxAmp}) the involved variables are equal to
\setcounter{equation}{28}
\begin{align}\label{NewNumChar}\nonumber
&\hat{\mu}_1=\sigma_1^2/a_1^2+1,\quad \hat{\sigma}_1^2=\sigma_1^4/a_1^4+2\sigma_1^2/a_1^2,\\
&\hat{\mu}_v=\sigma_v^2/a_v^2+1,\quad \hat{\sigma}_v^2=\sigma_v^4/a_v^4+2\sigma_v^2/a_v^2,
\end{align}
and
\begin{align} \label{hatRho}
\hat{\rho}_{v,1}=[&1+\sigma_1^2/a_1^2+\sigma_v^2/a_v^2+ \nonumber\\
&2\rho_{v,1}\sigma_1\sigma_v/a_1/a_v\cos(\varphi_1-\varphi_v)+\nonumber\\
&\sigma_1^2\sigma_v^2(0.5+0.625\rho_{v,1}^2)/a_1^2/a_v^2-\hat{\mu}_1\hat{\mu}_v]/\hat{\sigma}_1/\hat{\sigma}_v.
\end{align}

\begin{IEEEproof}
See the Appendix.
\end{IEEEproof}

\subsection{RMSE of OMA-Based Calibration}\label{S4.B}

Using (\ref{ApproxAmp}) and (\ref{ApproxPhs}) to analyze the \mbox{performance of OMA} assisted calibration is straightforward. The auto-correlation matrix of the noise term in (\ref{wOforthogonalScheme}) takes the form $\text{E}\left\{\boldsymbol{x}\boldsymbol{x}^\text{H}\right\}=\sigma^2\textbf{I}_V$, which implies in this case $\sigma_1^2=\sigma_v^2=\sigma^2$ and $\rho_{v,1}=0$, for all $2\leq v\leq V$. Before invoking (\ref{ApproxAmp}) and (\ref{ApproxPhs}) we also need the SNR at the DMFs' output, i.e. $\{a_v^2/\sigma_v^2|$ $1\leq v\leq V\}$. By the definition of MF, the instantaneous SNR of the correlation peak yielded by the $v$-th DMF amounts to
\begin{equation} \label{SNRdmf}
\frac{a_v^2}{\sigma^2}=\frac{E_v}{N_0},
\end{equation}
where $N_0$ is the single-sided noise PSD at the ground station, and $E_v$ is the energy of the calibrating waveform of element $v$, seen by the ground station within the period of $[0,T_\text{s})$. To be more specific, if the calibrating signal transmitted by element $v$ bears a power of $S_v$ (Watts) as it reaches the ground station, then $E_v=S_vT_\text{s}$. To initialize the RMSE computation, we may choose an arbitrary positive value as $\sigma^2$, and then determine $a_v^2$ according to a preset $E_v/N_0$ and (\ref{SNRdmf}). From a link budget perspective, $E_v/N_0$ is dependent on several key parameters in the way as follows
\begin{equation} \label{LinkBudget}
\frac{E_v}{N_0}=\frac{\text{EIRP}_v}{L_\text{p}}\left(\frac{G_\text{r}}{T_\text{e}}\right)\frac{T_\text{s}}{k_\text{B}},
\end{equation}
where $\text{EIRP}_v$ denotes the Equivalent Isotropic Radiating Power of element $v$, $L_\text{p}$ stands for the propagation loss, $k_\text{B}=-228$ $\text{dBw}\text{/Hz/K}$ represents the Boltzman constant. As the quotient of the receiving antenna's gain $G_\text{r}$ divided by the equivalent noise temperature $T_\text{e}$, $\left(G_\text{r}/T_\text{e}\right)$ is typically known as the \emph{quality factor} of the ground station.

\subsection{RMSE of CSmS-NOMA Based Calibration}\label{S4.C}

It is worth highlighting that in CSmS-NOMA the noise components in the estimated CCGs are no longer independent. Determining the correlation coefficient defined by (\ref{rho}) will play a critical role in applying (\ref{ApproxAmp}) and (\ref{ApproxPhs}) to CSmS-NOMA-based calibration.  Let us reexamine (\ref{DC}), and denote the noise part in $\tilde{\boldsymbol{w}}$ and $\boldsymbol{p}$ \mbox{as $\boldsymbol{x}$ and $\boldsymbol{x}_{\boldsymbol{p}}$,} respectively. The auto-correlation matrix of $\boldsymbol{x}$ is given by
\begin{equation} \label{Rx}
\boldsymbol{R}_{\boldsymbol{x}}=\text{E}\{\boldsymbol{x}\boldsymbol{x}^\text{H}\}=\boldsymbol{M}^{-1}\boldsymbol{R}_{\boldsymbol{x}_{\boldsymbol{p}}}\boldsymbol{M}^{-1}. \end{equation}
where $\boldsymbol{R}_{\boldsymbol{x}_{\boldsymbol{p}}}=\text{E}\{\boldsymbol{x}_{\boldsymbol{p}}\boldsymbol{x}^\text{H}_{\boldsymbol{p}}\}\in\mathbb{C}^{V\times V}$ is conjugate symmetric. Let $\boldsymbol{R}_{\boldsymbol{x}_{\boldsymbol{p}}}(y,z)$ be the $(y,z)$-th entry of $\boldsymbol{R}_{\boldsymbol{x}_{\boldsymbol{p}}}$, by recalling the derivation of (\ref{pv}) and (\ref{peak}), we have
\begin{align} \label{Rxp}
&\boldsymbol{R}_{\boldsymbol{x}_{\boldsymbol{p}}}(y,z)= \nonumber\\
&\boldsymbol{m}^\text{T}\text{E}\left\{\left[\begin{array}{c}
      n_\text{B}(y-1) \\
      n_\text{B}(y)\\
      \vdots\\
      n_\text{B}(y+L-2)\\
    \end{array}\right]\left[\begin{array}{c}
      n_\text{B}(z-1) \\
      n_\text{B}(z)\\
      \vdots\\
      n_\text{B}(z+L-2)\\
    \end{array}\right]^\text{T}\right\}\boldsymbol{m}.
\end{align}

To further simplify (\ref{Rxp}), we assume $y\leq z$ (The other half entries of the matrix $\{\boldsymbol{R}_{\boldsymbol{x}_{\boldsymbol{p}}}(y,z)|$ $y>z\}$ could be obtained via the symmetry of $\boldsymbol{R}_{\boldsymbol{x}_{\boldsymbol{p}}}$). Given $1\leq y\leq z\leq V$, we have
\begin{equation} \label{Rxp2}
\boldsymbol{R}_{\boldsymbol{x}_{\boldsymbol{p}}}(y,z)=\boldsymbol{m}^\text{T}\left[\begin{array}{cc}
      \boldsymbol{A} & \boldsymbol{B} \\
      \boldsymbol{C} & \boldsymbol{A}^\text{T}\\
    \end{array}\right]\boldsymbol{m},
\end{equation}
where $\boldsymbol{A}=\textbf{O}_{(z-y)\times(L-z+y)}$, $\boldsymbol{B}=\textbf{O}_{(z-y)\times(z-y)}$ and $\boldsymbol{C}$ $=\sigma^2\textbf{I}_{(L-z+y)}$.  In finding (\ref{Rxp2}) we have exploited the fact that $\{n_{\text{B}}(k)|k=0,1,\cdots\}$ are i.i.d. zero mean Gaussian RVs with variance of $\sigma^2$. Once $\boldsymbol{R}_{\boldsymbol{x}_{\boldsymbol{p}}}$ is achieved, $\boldsymbol{R}_{\boldsymbol{x}}$ can be computed immediately based on (\ref{Rx}). The diagonal entries of $\boldsymbol{R}_{\boldsymbol{x}}$ are $\left\{\sigma^2_v|1\leq v\leq V\right\}$, and the second to  the $v$-th entries in the first column are $\{\text{E}\left(n_vn_1^\ast\right)|1\leq v\leq V\}$, thus all the ingredients we need to compute (\ref{rho}) are ready.

\section{Hardware Platform}\label{Hardware}

\begin{figure}\centering
\includegraphics[width=86mm]{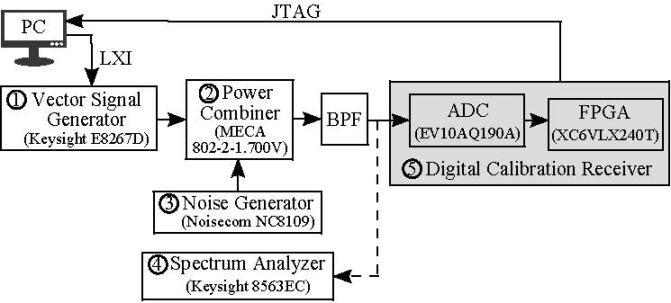}
\caption{Topology of the hardware platform.}
\label{Topology}
\end{figure}

\begin{figure}\centering
\includegraphics[width=86mm]{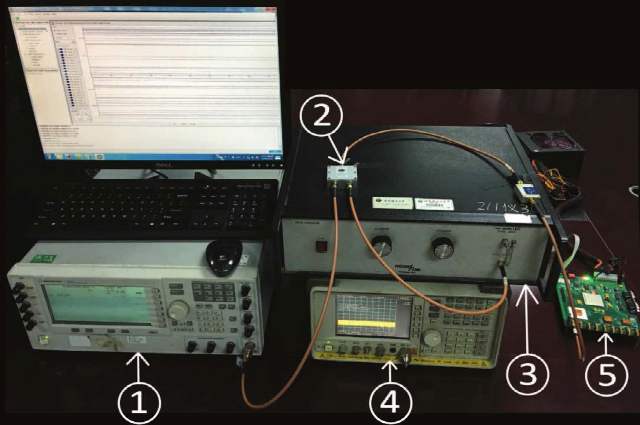}
\caption{Photograph of the hardware platform.}
\label{Photograph}
\end{figure}

In order to verify the effectiveness of CSmS-NOMA and confirm our theoretical conclusions derived above, we have built a hardware platform for real world test. The diagram and the photograph of the hardware platform are given in Fig.\ref{Topology} and Fig.\ref{Photograph}, respectively. The Keysight (a.k.a. Agilent) vector signal generator E8267D accepts a composite I/Q baseband discrete calibrating waveform from the PC, via the LAN eXtension for Instrument (LXI) protocol. Analog I/Q calibrating waveforms are restored inside E8267D by a dual-channel DAC, then passed through a quadrature modulator to produce the intermediate frequency (IF) calibrating signal. The IF is set to 10.7MHz to make best use of the components at hand. Without loss of generality, we set the chip rate of the DSSS calibrating signal at a relatively low level of $100$kHz, to fit in for the testing calibration receiver with limited computational capacity.

The composite I/Q waveform downloaded to E8267D is generated by a Matlab program running on a PC. The program first produces the DSSS calibrating waveform corresponding to each element with normalized amplitude, and then it performs a weighted sum according to a group of preset CCG values. In principle, this method is capable of accommodating arbitrary amount of elements with a single 8267D. It also guarantees an accurate amplitude \& phase relationship (as specified by the CCGs) at the input of the calibration receiver, as the common path from E8267D to the calibration receiver impacts all the individual calibrating signals in a non-discriminatory fashion.

To examine the influence of noise, the IF signal is combined with the output of a wideband (0Hz-300MHz) noise generator. The ADC samples at a rate of $40$MHz, hence to avoid aliasing we insert before the ADC input a band-pass filter (BPF) with center frequency of $10.7$MHz and passband width of $2$MHz. Another alternative direction of the BPF output to a spectrum analyzer (denoted by dash line in Fig.\ref{Topology}) is for measuring the power of the calibrating signal and the in-band noise, so we can determine the SNR in the experiment.

At the core part of the digital calibration receiver is an FPGA subsystem, on which we have \mbox{developed two} programs in Verilog Hardware Description Language (HDL). One is for OMA-based parallel DMFs, and the other is for CSmS-NOMA-based single DMF. Both of them may yield element-specific, complex-valued correlation peaks. With the aid of ChipScope (a Xilinx online diagnosis software), the peak values are uploaded to the PC via a Joint Test Action Group (JTAG) cable. Further processing is finished by the software Matlab. For OMA-based calibration, the estimates of relative amplitude \& phase among different elements may be extracted directly from the peak values. While for CSmS-NOMA, an extra ZF equalizer is required. With the calibration results and their preset true values at hand, we may compute empirical RMSEs following (\ref{AmpRMSE}) and (\ref{PhsRMSE}), by replacing the mathematical expectation therein with arithmetic average. To obtain each empirical result reported in the next section, $10^4$ independent experiments are carried out.

\section{Calibration Accuracy Results}\label{S6}

In this section, we investigate the achievable calibration accuracy by theoretical analysis, computer simulations and hardware experiment, respectively. The OMA-based strategies (using Walsh codes) and the proposed CSmS-NOMA approach are considered in a comparative way. Without loss of generality, all the elements are assumed to transmit at the same power, and their uncalibrated phases are assumed to be randomly distributed over the range of $[0,2\pi)$. The RMSEs of relative amplitude and phase estimates are chosen as the calibration accuracy metrics. In order to reveal the overall performance, in each simulation figures the RMSE has been averaged over all elements.

\setcounter{equation}{35}   %
\begin{figure*}[hb]
\hrulefill
\begin{align}
\label{app_1}
G_{v,1}^\text{RMSE}=
\sqrt{\left\{\text{E}\left[10\log_{10}\left(\frac{|\tilde{w}_v|/a_v}{|\tilde{w}_1|/a_1}\right)^2\right]\right\}^2+\text{var}\left[10\log_{10}\left(\frac{|\tilde{w}_v|/a_v}{|\tilde{w}_1|/a_1}\right)^2\right]}.
\end{align}
\end{figure*}
\setcounter{equation}{36}

\begin{figure}\centering
\includegraphics[height=76mm,width=82mm]{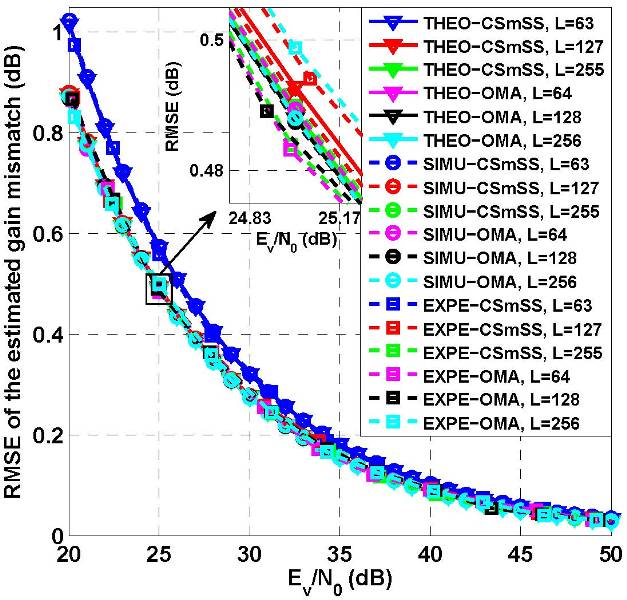}
\caption{RMSE of the gain mismatch estimates v.s. SNR, with fixed number of elements ($V=50$) and various signature code length $L$.}
\label{ResultsGainScenarioI}
\end{figure}

\begin{figure}\centering
\includegraphics[height=76mm,width=82mm]{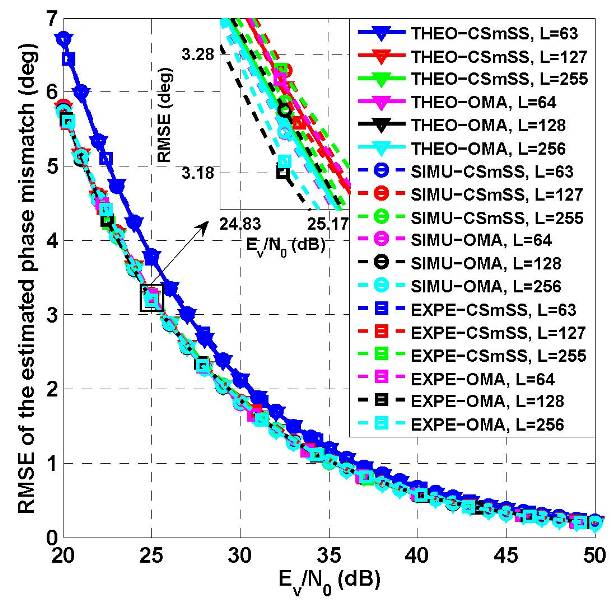}
\caption{RMSE of the phase mismatch estimates v.s. SNR, with fixed number of elements ($V=50$) and various signature code length $L$.}
\label{ResultsPhaseScenarioI}
\end{figure}

In Figs. \ref{ResultsGainScenarioI} and \ref{ResultsPhaseScenarioI} the calibration accuracy against the MF output SNR\footnote{As suggested by (\ref{LinkBudget}), when all the elements radiate at the same power, they will exhibit the same $E_v/N_0$ at the calibration receiver.} $E_v/N_0$, is shown with the number of elements being $50$ while the length of signature codes varying among $\{64,128,256\}$ for OMA, and $\{63,127,255\}$ for CSmS-NOMA. It is observed the accuracy of both amplitude and phase calibration improves monotonically as SNR increases. For all combinations of parameters, the analytical, the simulated and the experimental data manifest perfect accordance. Thanks to the orthogonality of signature codes, the precision of the OMA-based scheme is not affected by the code length, as long as the code length is larger than the \mbox{number of elements,} i.e. $V\leq L$. For the presented CSmS-NOMA approach, it is observed if the number of elements is close to the code length (e.g. $50$ v.s. $63$), the calibration precision suffers because of the Noise Enlarging Effect (NEE) of ZF, as mentioned in Section IV. The $E_v/N_0$ loss induced by NEE is about $1.5$dB in the low SNR, and it becomes less evident within the high-SNR range. Moreover, no perceivable loss is observed when the length of m-sequence increases to $127$ or $255$.


\begin{figure}[ht]\centering
\includegraphics[height=77mm,width=82mm]{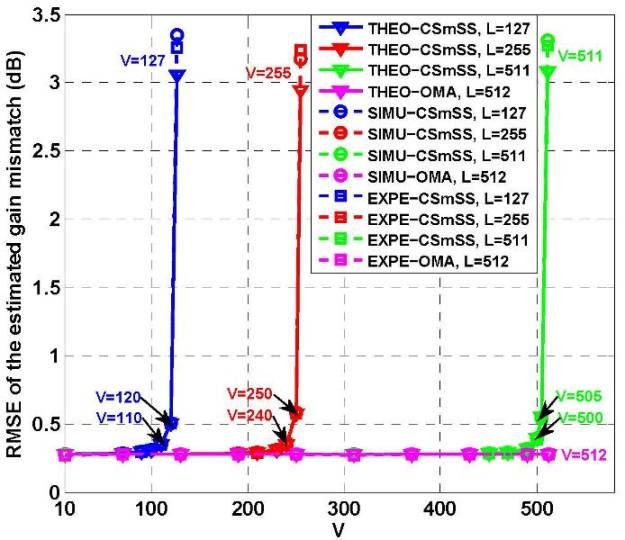}
\caption{RMSE of the gain mismatch estimates v.s. the number of elemental antenna ($V$) for various signature code length ($L$), with the SNR fixed at the level of $\frac{E_v}{N_0}=30\text{dB}$.}
\label{ResultsGainScenarioII}
\end{figure}
\begin{figure}[ht]\centering
\includegraphics[height=77mm,width=82mm]{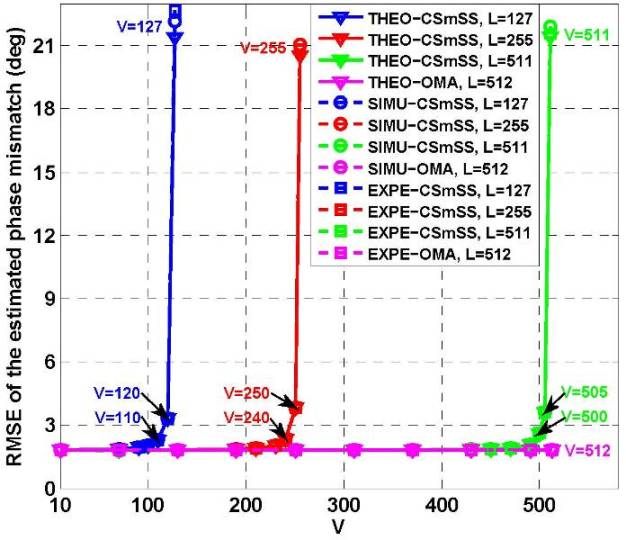}
\caption{RMSE of the phase mismatch estimates v.s. the number of elemental antenna ($V$) for various signature code length ($L$), with the SNR fixed at the level of $\frac{E_v}{N_0}=30\text{dB}$.}
\label{ResultsPhaseScenarioII}
\end{figure}

Figs. \ref{ResultsGainScenarioII} and \ref{ResultsPhaseScenarioII} are devoted to a further examination of the proposed CSmS-NOMA, which highlights the NEE-caused accuracy loss as the number of elements ($V$) approaches the code length ($L$). Again, it is witnessed the analytical, the simulated and the experimental results collaborate with one and another perfectly. In Figs.\ref{ResultsGainScenarioII} and \ref{ResultsPhaseScenarioII} we have considered m-sequences of three different lengths, namely $127$, $255$ and $511$. Moreover, the  performance  of the OMA-based strategy (512-chip Walsh code) is provided as a benchmark. Note that we omit the curves for $128$ and $256$-chip Walsh code, as the performance of OMA-based strategy is not dependent on the number of the elements, only if when $V\leq L$. It is found as $V$ approaches $L$, and especially in the extreme case where $V=L$, the accuracy of CSmS-NOMA may degrade to a considerable extend. It is also found the accuracy degradation of CSmS-NOMA occurs only when $V$ is close to $L$, e.g. when $V=110$ for $L=127$, when $V=240$ for $L=255$, or when $V=500$ for $L=511$. In other words, if the length of the m-sequence is chosen as properly larger than the number of elements, then CSmS-NOMA may perform as well as the OMA-based calibration.

\section{Conclusions}
NOMA-based parallel calibrations for large-scale spaceborne antenna arrays are investigated in this paper. Following the idea of NOMA, NOMA-based calibrating baseband sequences also named as CSmS-NOMA signaling are proposed, which is very suitable for large-scale spaceborne antenna array calibrations. The proposed scheme is easy to implement and enjoys satisfied performance.
It can strike tradeoffs between calibration complexity and performance.
Specifically, at cost of negligible performance loss resulting from non-orthogonality, the efficiency of calibrations can be significantly improved. Furthermore, the calibration performance of the large-scale antenna array calibrations has be analyzed theoretically. The proposed algorithm was also run on a hardware platform. Hardware platforms always involve some unpredict and uncontrollable factors introduced by hardware imperfections, which may prohibit theoretical algorithms' implementations. We reveals that our proposed algorithm is robust enough and can achieve the desired performance on the hardware platform.

\begin{figure*}[hb]
\hrulefill
\setcounter{equation}{39}
\begin{align}\label{PhsRMSE2}
P_{v,1}^\text{RMSE}\nonumber=&\frac{180}{\pi}\sqrt{\text{var}
\left[\left(\arg(\tilde{w}_v)-\arg(\tilde{w}_1)\right)
-\left(\varphi_v-\varphi_1\right)\right]
+\left\{\text{E}\left[\left(\arg(\tilde{w}_v)
-\arg(\tilde{w}_1)\right)-
\left(\varphi_v-\varphi_1\right)\right]\right\}^2}\nonumber \\
=&\frac{180}{\pi}\sqrt{\text{var}\left[\left(\arg(\tilde{w}_v)
-\varphi_v\right)-\left(\arg(\tilde{w}_1)-\varphi_1\right)\right]}.
\end{align}

\setcounter{equation}{45}
\begin{align}
\text{cov}\left[\left(\arg(\tilde{w}_v)-\varphi_v\right),
\left(\arg(\tilde{w}_1)-\varphi_1\right)\right]
=&\,\text{E}\left[\left(\arg(\tilde{w}_v)-\varphi_v\right)\left
(\arg(\tilde{w}_1)-\varphi_1\right)\right] \nonumber \\
\approx&\,\text{E}\left[\frac{\left(n_{\text{Q},v}\cos\varphi_v-
n_{\text{I},v}\sin\varphi_v\right)\left(n_{\text{Q},1}
\cos\varphi_1-n_{\text{I},1}\sin\varphi_1\right)}
{\left(a_v+n_{\text{I},v}\cos\varphi_v+n_{\text{Q},v}
\sin\varphi_v\right)\left(a_1+n_{\text{I},1}\cos\varphi_1
+n_{\text{Q},1}\sin\varphi_1\right)}\right].\label{PhsCov2}
\end{align}
\end{figure*}

\appendix[Derivation of $G_{v,1}^\text{RMSE}$ and $P_{v,1}^\text{RMSE}$]

In the appendix, the derivations for (\ref{ApproxAmp}) and (\ref{ApproxPhs}) are discussed in detail. At the beginning, (\ref{AmpRMSE}) is rewritten into an alternative form presented at the bottom of this page as (\ref{app_1}). As mentioned above, a prerequisite for (\ref{ApproxAmp}) and (\ref{ApproxPhs}) to hold is the calibration receiver has sufficiently high SNR. In this case, it is reasonable to expect the estimated gain mismatch approaches its true value, i.e., $\left(\frac{|\tilde{w}_v|/a_v}{|\tilde{w}_1|/a_1}\right)^2\rightarrow1$. Since $\underset{x\rightarrow1}{\lim}(\ln x)=x-1$, we have
\setcounter{equation}{36}
\begin{equation} \label{Appendix2}
\text{E}\left[10\log_{10}\left(\frac{|\tilde{w}_v|/a_v}{|\tilde{w}_1|/a_1}\right)^2\right]\approx\frac{10}{\ln(10)}\left\{\text{E}\left[\left(\frac{|\tilde{w}_v|/a_v}{|\tilde{w}_1|/a_1}\right)^2\right]-1\right\},
\end{equation}
and similarly
\begin{equation} \label{Appendix3}
\text{var}\left[10\log_{10}\left(\frac{|\tilde{w}_v|/a_v}{|\tilde{w}_1|/a_1}\right)^2\right]\approx\left[\frac{10}{\ln(10)}\right]^2\cdot\text{var}\left[\left(\frac{|\tilde{w}_v|/a_v}{|\tilde{w}_1|/a_1}\right)^2\right]. \end{equation}

To further simplify the mathematical formulation the statistical characteristics of $\left(\frac{|\tilde{w}_v|/a_v}{|\tilde{w}_1|/a_1}\right)^2$ will be further exploited. As the squared sum of two independent non-zero mean normally distributed Random Variables (RVs), $2|\tilde{w}_v|^2/\sigma_v^2$  obeys non-central Chi-square distribution, with its degree of freedom and its non-central parameter (NCP) being $2$ and $2a_v^2/\sigma_v^2$, respectively. Given the ``high SNR" assumption, we have $2a_v^2/\sigma_v^2\gg1$, hence the non-central Chi-square distribution can be approximated by Gaussian distribution \cite{Patnaik1949}, i.e. $(|\tilde{w}_v|/a_v)^2\sim\mathcal{N}\left(\hat{\mu}_v,\hat{\sigma}_v^2\right)$,
and $(|\tilde{w}_1|/a_1)^2\sim\mathcal{N}\left(\hat{\mu}_1,\hat{\sigma}_1^2\right)$,
where $\hat{\mu}_v$, $\hat{\mu}_1$, $\hat{\sigma}_v$ and $\hat{\sigma}_1$ are defined in (\ref{NewNumChar}).

The ratio of two normally distributed RVs (not necessarily independent), including its probability density function (PDF) and numerical characteristics, have been thoroughly studied in \cite{PhamGia2007} and \cite{Hayya1975}. Quoting the conclusions therein, we arrive at
\begin{subequations} \label{AppendixEq1}
\begin{align}
&\text{E}\left[\left(\frac{|\tilde{w}_v|/a_v}{|\tilde{w}_1|/a_1}\right)^2\right]\approx\frac{\hat{\mu}_v}{\hat{\mu}_1}+\frac{\hat{\sigma}_1^2\hat{\mu}_v}{\hat{\mu}_1^3}-\frac{\hat{\rho}_{v,1}\hat{\sigma}_1\hat{\sigma}_v}{\hat{\mu}_1^2},\\
&\text{var}\left[\left(\frac{|\tilde{w}_v|/a_v}{|\tilde{w}_1|/a_1}\right)^2\right]\approx\frac{\hat{\sigma}_1^2\hat{\mu}_v^2}{\hat{\mu}_1^4}+\frac{\hat{\sigma}_v^2}{\hat{\mu}_1^2}-\frac{2\hat{\rho}_{v,1}\hat{\sigma}_1\hat{\sigma}_v\hat{\mu}_v}{\hat{\mu}_1^3},
\end{align}
\end{subequations}
where $\hat{\rho}_{v,1}$ is the correlation coefficient between $|\tilde{w}_v|^2/a_v^2$ and $|\tilde{w}_1|^2/a_1^2$. Note (\ref{AppendixEq1}) holds on the conditions $\hat{\mu}_1\gg\hat{\sigma}_1$ and $\hat{\mu}_v\gg\hat{\sigma}_v$ \cite{PhamGia2007},\cite{Hayya1975}. Recalling (\ref{NewNumChar}), one may find these conditions are met as $2a_v^2/\sigma_v^2\gg1$. According to its definition, $\hat{\rho}_{v,1}$ in (\ref{AppendixEq1}) may be computed by the following formula
\begin{equation}\nonumber
\hat{\rho}_{v,1}=\frac{\text{E}[(|\tilde{w}_v|/a_v)^2(|\tilde{w}_1|/a_1)^2]-\hat{\mu}_1\hat{\mu}_v}{\hat{\sigma}_v\hat{\sigma}_1}.
\end{equation}

After a tedious but straightforward derivation, the above formula may eventually be simplified into (\ref{hatRho}). In that process we need to invoke a theorem on the variance of the product of two correlated Gaussian RVs \cite{Ware2003}, which reads as:
\emph{``For two RVs that obey} $\mathcal{N}\left(\mu_1,\sigma_1^2\right)$ \emph{and} $\mathcal{N}\left(\mu_2,\sigma_2^2\right)$ \emph{respectively, their product $Y$ has the variance}
\begin{align} \nonumber
V(Y)=\mu_1^2\sigma_2^2+\mu_2^2\sigma_1^2+\sigma_1^2\sigma_2^2+2\rho\mu_1\mu_2\sigma_1\sigma_2+\rho^2\sigma_1^2\sigma_2^2,
\end{align}
\emph{where $\rho$ is the correlation coefficient between the two Gaussian distributed RVs."}Substituting (\ref{AppendixEq1}) into (\ref{Appendix2})-(\ref{Appendix3}), and then (\ref{Appendix2})-(\ref{Appendix3}) into (\ref{app_1}), we finally reach (\ref{ApproxAmp}), the approximate formula of $G_{v,1}^\text{RMSE}$. Now we proceed with the approximate RMSE of the phase mismatch estimate, namely (\ref{ApproxPhs}). Again, let us rewrite the definition of $P_{v,1}^\text{RMSE}$ (\ref{PhsRMSE}) into an equivalent form, which is shown at the bottom of this page as (\ref{PhsRMSE2}).

To further simplify (\ref{PhsRMSE2}), we need to evaluate the following formula
\setcounter{equation}{40}
\begin{align}\label{VarPhs}\nonumber
&\text{var}\left[\left(\arg(\tilde{w}_v)-\arg(\tilde{w}_1)\right)-\left(\varphi_v-\varphi_1\right)\right]\\\nonumber
&=\text{var}(\arg(\tilde{w}_v)-\varphi_v)+\text{var}(\arg(\tilde{w}_1)-\varphi_1)\\
&\quad-2\text{cov}\left[\left(\arg(\tilde{w}_v)-\varphi_v\right),
\left(\arg(\tilde{w}_1)-\varphi_1\right)\right].
\end{align}
Recall the definition of $\tilde{w}_v$ in the \emph{Proposition}, we have
\begin{align}\label{ApproxPhsv}\nonumber
\arg(\tilde{w}_v)-\varphi_v&=\arctan\left(\frac{n_{\text{Q},v}\cos \varphi_v-n_{\text{I},v}\sin \varphi_v}{a_v+n_{\text{I},v}\cos \varphi_v+n_{\text{Q},v}\sin \varphi_v}\right)\\
&\approx\frac{n_{\text{Q},v}\cos \varphi_v-n_{\text{I},v}\sin \varphi_v}{a_v+n_{\text{I},v}\cos \varphi_v+n_{\text{Q},v}\sin \varphi_v}.
\end{align}
The approximation in the above formula will be tight when the SNR is large enough. In the high SNR case, $\arg(\tilde{w}_v)$ is close to its true value $\varphi_v$, and $\underset{x\rightarrow0}{\arctan}(x)\approx x$. For the term of $\frac{n_{\text{Q},v}\cos \varphi_v-n_{\text{I},v}\sin \varphi_v}{a_v+n_{\text{I},v}\cos \varphi_v+n_{\text{Q},v}\sin \varphi_v}$ both its numerator and denominator are Gaussian distributed, i.e.,
\begin{subequations}
\begin{align}
&\left(n_{\text{Q},v}\cos \varphi_v-n_{\text{I},v}\sin \varphi_v\right)\sim\mathcal{N}\left(0,\hat{\sigma}_v^2/2\right),\\
&\left(a_v+n_{\text{I},v}\cos \varphi_v+n_{\text{Q},v}\sin \varphi_v\right)\sim\mathcal{N}\left(a_v,\hat{\sigma}_v^2/2\right).
\end{align}
\end{subequations}
It is obvious that the correlation coefficient between these two Gaussian RVs is zero, and thus they are statistically independent. Furthermore, the statistical characteristics of the ratio between a Gaussian RV divided by another independent Gaussian RV have been given in \cite{Marsaglia2006}. Therefore, we have the following equalities
\begin{subequations}
\begin{align}
&\text{E}\left[\arg(\tilde{w}_v)-\varphi_v\right]=0, \\
&\text{var}\left[\arg(\tilde{w}_v)-\varphi_v\right]\approx\sigma_v^2/2a_v^2. \label{var-v}
\end{align}
\end{subequations}Following a similar logic, referring to $(\arg(\tilde{w}_1)-\varphi_1)$ the following equalities also hold
\begin{subequations}
\begin{align}
&\text{E}\left[\arg(\tilde{w}_1)-\varphi_1\right]=0, \\
&\text{var}\left[\arg(\tilde{w}_1)-\varphi_1\right]\approx\sigma_1^2/2a_1^2. \label{var-1}
\end{align}
\end{subequations}

Besides (\ref{var-v}) and (\ref{var-1}), the remaining term in  (\ref{VarPhs}) we still need to compute is the covariance between $\arg(\tilde{w}_v)-\varphi_v$ and $\arg(\tilde{w}_1)-\varphi_1$.  According to its definition, the covariance in (\ref{VarPhs}) may be calculated by the first equation as (\ref{PhsCov2}) at the bottom of this page.
Plugging (\ref{ApproxPhsv}) into the first equation of (\ref{PhsCov2}), the final equation can be achieved, which takes the form as the expectation of two RVs' ratio. In \cite{Casella2002}, approximate formula for expectation of this kind has been studied. Invoking the conclusions therein [30.\,Section 5.5.4] for (\ref{PhsCov2}) leads to
\setcounter{equation}{46}
\begin{align}\label{PhsCovApprox}\nonumber
&\text{cov}\left[\left(\arg(\tilde{w}_v)-\varphi_v\right),\left(\arg(\tilde{w}_1)-\varphi_1\right)\right]\\
&\qquad\approx\frac{\rho_{v,1}\sigma_1\sigma_v\cos(\varphi_1-\varphi_v)}{2a_va_1+\rho_{v,1}\sigma_1\sigma_v\cos(\varphi_1-\varphi_v)}.
\end{align}
Substituting (\ref{var-v}), (\ref{var-1}) and (\ref{PhsCovApprox}) into (\ref{VarPhs}), and then putting (\ref{VarPhs}) into (\ref{PhsRMSE2}), finally (\ref{ApproxPhs}) can be achieved. Thereby the proof for the approximate formula of $P_{v,1}^\text{RMSE}$ has been finished.



\end{document}